**Classification:** Physical Sciences: Earth, Atmospheric and Planetary Sciences.

# Title: Precursory moment release scales with earthquake magnitude.

**Authors:** M. Acosta[1]*, F.X. Passelègue[1], A. Schubnel[2], R.Madariaga[2], M. Violay[1]

**Affiliations:**

[1]EPFL, LEMR, Lausanne, Switzerland,

[2]Laboratoire de Géologie, CNRS, UMR, ENS, Paris, FR.

*Correspondence to: mateo.acosta@epfl.ch. ORCID#0000-0002-0098-7912.

**Abstract:**

Today, observations of earthquake precursors remain widely debated. While precursory slow slip is an important feature of earthquake nucleation, foreshock sequences are not always observed and their temporal evolution remains unconstrained. Here, we report on stick-slip experiments (laboratory earthquakes) conducted under seismogenic stresses in dry and fluid pressure conditions. We show that the precursory moment release scales with mainshock magnitude irrespective of the slip behavior (seismic or aseismic), the presence of fluid and the fault's slip history. Importantly, this observation is supported by earthquake nucleation theory and holds for natural earthquakes in a magnitude range from $M_w$6.0 to $M_w$9.0. Even though a large gap remains between laboratory and natural observations, moderate to large earthquakes may be foresighted through integrated seismological and geodetic measurements of both seismic and aseismic slip during earthquake nucleation.

**Significance Statement:**

Understanding the slow nucleation phase preceding dynamic earthquake ruptures is crucial to assess seismic hazard. Here, we show that the temporal evolution of laboratory earthquake precursors (precursory slow slip, fault coupling and precursory seismicity) are of little use to assess seismic hazard. Nevertheless, irrespective of fault slip behavior (seismic or aseismic) and environmental conditions (stress state; fluid pressure and slip history), the amount of moment released during the precursory phase scales with earthquake magnitude. This property is demonstrated by laboratory observations, earthquake nucleation theory and, by natural observations of earthquakes ranging from $M_w$6.0 to $M_w$9.0. Larger earthquakes must therefore exhibit a larger nucleation phase and consequently are more likely to be detectable.

**Main Text:**

Understanding earthquake initiation is crucial to assessing seismic hazard. Theoretical and numerical models using either slip-weakening laws (*1-3*) or rate and state friction laws (*4-7*) predict that earthquake rupture is preceded by a nucleation phase (NP). The rupture accelerates over a growing fault patch until it reaches a characteristic length $L_c$, at which dynamic rupture propagation initiates. While the NP has been studied in the laboratory (*2, 8-13*), natural observations of such NP are scarce and remain debated (*14-17*). The study of recent well-instrumented moderate and large magnitude earthquakes has highlighted that several ruptures were preceded by precursory slip, linked or not with foreshock sequences (*16-21*). However, other observations indicate that, rather than precursory slip, foreshocks trigger one another in a cascade-like manner up to the main rupture (*15*). This debate remains unresolved mainly because the temporal evolution of both slip and seismicity and their intensity compared to the mainshock remain poorly constrained. The answers to the above questions might be found at the laboratory scale, where rupture propagation and fault conditions are well-controlled.

Here, we carefully analyzed the NP of 150+ laboratory stick-slip events (proxies for earthquakes) conducted under stress conditions representative of the upper seismogenic crust, under both dry and pressurized fluid conditions. We conducted experiments at confining pressures ($\sigma_3$) ranging between 50 to 95 MPa and pore fluid pressures ($p_f$) from 0 to 60 MPa. In each case, the displacement-stress curve (Fig.1a,c., *Methods*) revealed an initial elastic loading phase until a stress state $\tau_s$. Then, the fault started to slip in a stable manner (*2, 10, 11*). When precursory slip ($u_{prec}$) finished, the fault's maximum shear strength ($\sigma_0$) was reached and the instability propagated with fast co-seismic slip ($u_{cos}$) and a sudden stress drop ($\Delta\sigma_{cos}$) down to a final stress value ($\sigma_f$). When co-seismic slip arrested, as constant loading continued, a new elastic loading phase ensued, repeating the stick-slip cycle (Fig.1).

Under dry conditions (Fig.1a,b), the NP is characterized by an exponential slip acceleration following $u_{prec}(t)=u_0*\exp((t_o-t)/t_c)$ where $t_0$ is the time of the mainshock, $t_c$ the characteristic nucleation time (9), and $u_0$ is fault slip at the beginning of the acceleration phase. Note that both temporal constants can be estimated only after the mainshock occurrence. Analysis of acoustic emissions activity (AEs) in our experiments (*Methods*) highlights that, under dry conditions, AEs also followed an exponential increase until ($t_0$) (Fig.1b insert), consistent with the exponential acceleration of $u_{prec}$ (13). Subsequent stick-slip events led to an increase of the total AEs (from 31 in cycle #1 to 79 in cycle #10). With pressurized fluids (Fig.1c, d), the NP is characterized by two clearly distinguishable slip acceleration stages. The first one is described by an exponential acceleration as observed in dry conditions. However, a transition from an exponential growth to a power law is observed close to dynamic rupture propagation. This change in acceleration of $u_{prec}$ is comparable to the nucleation process observed through high-speed photo-elasticity in transparent polymers (9). There, the length of the ruptured area within the nucleation zone patch grew first exponentially with time, and, once the size of the rupture patch became comparable to that of the characteristic nucleation length, the growth followed as an inverse power law of time. This second slip acceleration stage can be very short during dry experiments; potentially the reason why it was not observed. Surprisingly, with pressurized fluids, $u_{prec}$ was not associated with any AEs. It is unlikely that the absence of foreshocks in all fluid-pressurized conditions resulted solely from increased attenuation of seismic waves since the sensors were located <1 mm away from the fault. Previous experiments have indeed shown that AEs are systematically observed prior to the failure of intact or thermally cracked, dry or water saturated granite specimens (*22*).

To further investigate the NP of laboratory earthquakes (Fig.2), we computed the temporal evolution of the fault's mechanical coupling (*FC*, *Methods*, Ref.*24*) as a function of time to mainshock during earthquake sequences recorded both in dry and water-pressurized conditions. As expected, the fault remained strong (*FC*~1) during elastic loading. At the onset of the mainshock, $FC(t_0)$ decreased to ~0.85- 0.7 due to initiation of the nucleation process in dry conditions. With subsequent events, $FC(t_0)$ decreased due to fault surface evolution (Fig2. darkest traces; Ref.*10*). With pressurized fluids, when nucleation initiated, *FC* consistently decreased to ~0.5 at $t_0$. No influence of the sliding history and fault surface evolution was noted. The reduced fault coupling and lack of foreshocks (or their size reduction) under fluid pressure conditions could be due to a local increase of the nucleation length at the scale of fault's asperities. The presence of local fluid overpressures influences the dynamics of fault's asperities by changing the distribution of frictional heterogeneities which in turn control both fault coupling and foreshock dynamics (*6,7*).

We now estimate the moment release of both the precursory and co-seismic stages (Fig.3a, Methods). While the temporal evolution of slip and seismicity during nucleation depends on the fault's state of stress (*13*), fluid pressure level and cumulative fault slip, we observe that in our experiments $M_p$ systematically scales with $M_0$ (i.e. the magnitude of the instability) (Fig.3a). The scaling can be shown combining earthquake nucleation theory with the scaling between earthquakes' fracture energy and co-seismic slip. In fact, experimental (*2,24-25*), seismological (*26*) and theoretical studies (*27*) have demonstrated that the fracture energy of earthquakes increases as a power law of their co-seismic slip following: $G = a\, u_{cos}^{\alpha}$, where *a* is a scaling pre-factor, and $\alpha$ a given power (in our experiments a~1.22e10 and $\alpha$~*1.783* (Fig.S4, *Methods*).

G can then be used in a small-scale-yielding description to estimate an upper-end value of the nucleation length (*1-3*) following: $L_c = \frac{2\mu G}{\Delta\sigma_d^2}$, with $\Delta\sigma_d$ the earthquake's dynamic stress drop. Then, using usual seismological relations (*Methods*), we get the following scaling relation between precursory and co-seismic moments:

$$M_p = \frac{(2a.\mu^{1-\alpha})^3}{C^{2\alpha}} \cdot \frac{\Delta\sigma_s^{2\alpha}}{\Delta\sigma_d^5} \cdot M_0^\alpha$$

*(Eq1)*

This relationship shows that the larger the released precursory moment, the larger the co-seismic moment of the earthquake. Taking common stress drop values in our experiments ($\Delta\sigma_s$ = 2 to 40 MPa and $\Delta\sigma_d$ = 5 to 90 MPa, Table.S1) we observe that this relation quantitatively predicts the precursory moment release observed in our experiments. Note that such $\Delta\sigma_s$ and $\Delta\sigma_d$ values are higher than usual earthquake stress drops, due to our finite experimental fault and fixed rupture area. In laboratory earthquakes, most of the elastic energy is accumulated in the apparatus column, i.e. within a volume considerably larger than the sample. The larger the normal stress ($\sigma_{N0}$) acting on the fault, the larger the elastic energy stored within the sample/apparatus medium and consequently, the larger the coseismic slip and fracture energy if compared to an infinite fault. Therefore, increasing $\sigma_{N0}$ in our experiments does not necessarily imply a reduction of $L_c$ (*2,4,6,9,13*) because of the larger $\Delta\sigma_s$ and $\Delta\sigma_d$. In our experiments, the scaling arises from the relation between $u_{prec}$ and $u_{cos}$ (Fig.S4) and it is noteworthy that $u_{prec}$, $u_{cos}$, $\Delta\sigma_s$, and $\Delta\sigma_d$ evolved spontaneously when faults reached the conditions for instability. Therefore, the scaling between $M_P$ and $M_0$ resulted from the final value of these four quantities, resulting in values for $\alpha$ slightly larger than those of natural observations (*25-26*). The scaling is also confirmed on mechanical data obtained by previous studies of laboratory earthquakes (Fig.3a, empty symbols, Methods), highlighting the independence from the experimental set-up.

While the precursory moment release remains undetermined in most natural and anthropogenic seismicity (*28*), several examples of well-instrumented earthquakes where the NP is properly characterized follow the same scaling relation as our laboratory earthquakes. This is especially valid when geodetic and/or repeater inferred measurements of precursory moment release are taken into account (Fig3.b, black squares; *Methods*, Refs:*17, 19-21, Supplementary References*). In addition, considering $M_p$ only from the release of seismic moment during foreshock sequences (i.e. only from seismic slip) results in large undershoot with respect to the proposed scaling (Fig3.b red star (*18*)). If the proposed scaling property is correct, this undershoot could occur because part of the precursory deformation is accommodated by aseismic slip transients (*17, 19-22*) or by seismic events of magnitude smaller than the magnitude of completeness of the catalogs. While this feature is clearly observed in laboratory experiments (Fig.1-3, Refs: *2, 8-13*), natural observations of aseismic slip are still rare and debated (*14-17*). In the light of our results and because current geodesy generally lacks the resolution to observe aseismic slip (*14, 17*), some earthquake sequences may nevertheless present a cascade-like initiation (i.e. small events trigger one another until the main rupture (*15*)). Yet, where transients of aseismic slip can be resolved, earthquakes seem to nucleate through a slow slip trigger (*16-17, 19-21*). When considering precursory aseismic slip, experimental and seismological observations suggest that $M_p$, i.e. the energy released during NP, increases with the seismic moment of the mainshock. In fact, applying *Eq1* with the scaling proposed by Abercrombie and Rice (Ref:26; ($\alpha$ ~1.28 and a= 5.25e6)) and usual (*29*) earthquake stress drops ($\Delta\sigma_s$~ 0.1-1 MPa and $\Delta\sigma_d$ =1-10 MPa) our scaling provides conservative estimates for natural earthquakes (Fig.3b, dotted lines).

Our study has strong implications for earthquake nucleation. First, because the precursory moment release systematically follows an exponential growth during the first acceleration phase, continuously monitoring the time evolution of fault coupling could provide first order information about the fault's stability, and the eventual rupture initiation. Second the occurrence of a large earthquake ($M_w$7 or higher) does not necessarily imply that a larger earthquake will not occur in the days following it. If the released seismic moment contributes to the precursory moment for a bigger asperity, a larger earthquake can occur in a close timescale, as was the case for the 2011 $M_w$9.0 Tohoku-Oki (*20*), the 2014 $M_w$8.1 Iquique (*21*) and the 1960 $M_w$9.4-9.6 Valdivia (*30*) earthquakes, all of which were preceded by large $M_w$ foreshocks. Third, and foremost, independent of initial conditions (fluid pressure, stress and slip history), the larger $M_p$, the larger $M_0$ (Fig. 3) whether it is released seismically, aseismically or by a combination of both. This confirms that both the number of foreshocks and their characteristic acceleration can differ for a given earthquake magnitude (*19*). Moreover, fluid pressures are likely to reduce fault coupling (*17,31*), compared to dry conditions, and, in light of our experiments, regulate foreshock sequences. Significantly, $M_p$ increases with mainshock magnitude. Therefore, the larger the fault patch which will rupture, the larger $L_c$ and the higher the possibility of detecting and following precursory activity of moderate to large earthquakes by combined geodetic and seismological methods.

**Materials and Methods**

**Starting samples**

Experimental samples consisted of Westerly Granite (WG) cylinders of 40 mm diameter and 88 mm length. WG is a material representative of the upper continental crust and suitable for laboratory work due to its low alteration, low anisotropy, homogeneity, fine grain size and simple mineralogy. The cylinders were initially heat-treated at 450°C in order increase their permeability by one order of magnitude (~$10^{-19}$ m$^2$) and consequently, to allow reasonable fluid diffusion and saturation times. Cylinders were then saw-cut at an angle ($\theta$) of 30º to the sample's long axis to create an artificial elliptical fault (~80 mm length and ~40 mm width). The faults' surfaces were grinded to ensure perfect contact and roughened with #240 grit paper in order to ensure a minimum cohesion along the fault's interface and impose a constant fault roughness in all the specimens.

**Triaxial apparatus and pore fluid system**

Experiments were performed on the oil-medium, tri-axial apparatus of ENS Paris. The apparatus is able to support 100 MPa in confining pressure and 680 MPa in axial stress for samples of 40 mm diameter. Both axial and confining stresses were servo-controlled independently and recorded with 1kPa resolution. Fluid pressures were servo-regulated through a *Quizix 20k* double syringe pump of 120 MPa maximum capacity (1 microliter and 1 kPa volume and pressure accuracy). Axial displacements ($u_{ax}$) were recorded through three gap sensors located outside the cell with a resolution of 0.1 μm at 100 samples per second.

**Acoustic emission monitoring**

During experiments, acoustic activity was monitored through 15 piezo-ceramic sensors which consist of a PZT crystal (PI ceramic PI255, 0.5 mm thick and of diameter 5 mm) contained in a brass casing. The sensors were glued directly on the samples with cyanoacrylate adhesive following the sensor map in Fig.S1. Acoustic waveforms were recorded with two different techniques (*13*). First, each unamplified signal was relayed to a digital oscilloscope allowing for the recording of macroscopic stick-slip events within a time window of 6.5 ms at 10 MHz (*13*). Second, to record low amplitude acoustic emissions activity, signals were amplified at

45 dB through pre-amplifiers. Amplified signals were then relayed to a trigger logic box. Using this second system, AE's were recorded if at least 4 sensors recorded an amplitude larger than a given threshold, that is set at 0.001 Volts. The complete waveform catalogue was then manually analyzed to remove possible triggers from background noise.

**High frequency stress and strain monitoring**

Four strain gauges shown in Fig.S1. were glued ~1 mm away from the fault and allowed a local recording of the axial ($\varepsilon_1$) and radial ($\varepsilon_3$) strains at 10 MHz sampling frequency. The gauges were wired in a full (Wheatstone) bridge configuration, allowing the direct measurement of $\varepsilon_1$-$\varepsilon_3$ through the 4*350 $\Omega$ resistors. To calibrate the gauges, we assumed a constant Young's modulus of the rock during the elastic loading phase of each event such that we had direct conversion from the strain recorded at the gauge to the corresponding far field differential stress ($\Delta\sigma$). This near-fault sensor allowed recording the dynamic character of each stick-slip event (*32*) and therefore to estimate the fracture energy of the event (*24*).

**Loading procedure**

For each test, axial and radial pressures (respectively ($\sigma_1$) and ($\sigma_3$)) were increased up to 10 MPa. Then, in the case of pore fluid experiments, air was cautiously flushed from the sample by increasing fluid pressure ($p_f$) at the lower end of the sample. Once fluid percolated through the entire sample, fluid pressure was increased up to 5 MPa at the upper and lower ends until pressure and volume equilibrium were reached at both ends. Experiments were then conducted at different effective pressure conditions, where $\sigma'=\sigma - p_f$ accounts for effective stress. Finally, axial stress was increased imposing a constant volume rate in the axial piston which resulted in initial strain rates- ranging from ~$1.10^{-5}$ s$^{-1}$ to ~$3.10^{-5}$ s$^{-1}$ while ($\sigma_3$) and ($p_f$) were held constant. Under our experimental configuration, the fault's shear stress ($\tau$) and effective normal stress ($\sigma_n$') were computed following:

$$\tau = \frac{\sigma_1' - \sigma_3'}{2} * \sin(2\theta)$$

*(Eq2)*

and

$$\sigma_n = \frac{\sigma_1' - \sigma_3'}{2} * (1 - \cos 2\theta) + \sigma_3'$$

*(Eq3)*

Due to our experimental configuration, shear and normal stresses increased simultaneously as $\sigma_1$' increased.

Fault displacement ($u_f$) was computed from the gap sensors located outside the pressure vessel by correcting the direct measurement of axial displacement ($u_{ax}$) from the stiffness of the experimental apparatus ($Ey_{col}$ ~38 GPa) and by projecting the displacement on the fault plane following:

$$u_f = \frac{u_{ax} - Ey_{col} * L * \Delta\sigma}{\cos\theta}$$

*(Eq4)*

where $L$ is the sample's length; and $\Delta\sigma$ is the deviatoric stress.

A summary of the 150+ recorded stick-slip events is presented in Supplementary Table 1. In these experiments, radial stresses ranged from 50 to 95 MPa and pore fluid pressures from 0 to 60 MPa.

**Mechanical coupling of experimental faults**

To evaluate the degree of mechanical coupling of the experimental fault during stick-slip cycles (*FC*), we first computed the fault slip rate ($\dot{u}_f$) recorded using the external gap sensors with 1 s centered time windows such that:

$$\dot{u}_f(t) = \frac{u(t+0.5\ s) - u(t-0.5\ s)}{((t+0.5\ s) - (t-0.5\ s))}$$

*(Eq5)*

Due to the fast strain rates imposed in our experiments compared to tectonic loading rates, we assumed that all deviations in strain rate from a fully coupled fault resulted from slip along the fault (*10*). Then, we defined the mechanical coupling of the fault as the ratio between the estimated fault slip rate to the imposed displacement rate when the fault is fully coupled ($\dot{u}_0$) such that: $FC = \left(\frac{\dot{u}_f}{\dot{u}_0}\right)$. This estimation of the fault coupling is comparable with the one derived from geodetic measurements along natural faults (*23, 33*).

**Corrections for elastic displacement and calculation of precursory and co-seismic moment in laboratory earthquakes.**

In all our experiments (Table.S1), the corrections for elastic displacement of the sample and apparatus deformation were performed replacing $Ey_{col}$ in (*Eq4*). by a new constant $Ey_{system}$ which was computed individually for each event. In that sense, the change in elasticity was corrected after each dynamic slip event. For the external points (Refs. *34-36* the data was manually recovered. Then, the data were corrected for elasticity and the shear stress –slip curves were plotted (examples are shown in Fig.S2.). Values of maximum shear stress at the onset of instability ($\tau_0$), shear stress drop ($\Delta\sigma$), precursory slip ($u_{prec}$) and co-seismic slip ($u_{cos}$) were recovered from those curves (Fig. S2.) and reported in Table.S2. From that table, the precursory and co-seismic moments were computed as $M_p = \mu.A.u_{prec}$ and $M_0 = \mu.A.u_{cos}$ respectively taking an average rock shear modulus of $\mu$=30 GPa (Fig. S3.)

**Precursory and co-seismic moment release: Theory.**

From a theoretical viewpoint, assuming that NP ends with a linear slip weakening law similar to dynamic rupture, the precursory moment can be written as $M_p = \mu.D_c.L_c^2$, where $D_c$ is the characteristic slip weakening distance. The earthquake's fracture energy is $G = \frac{\Delta\sigma_d D_c}{2}$ with $\Delta\sigma_d$ the earthquake's dynamic stress drop. G can then be used in a small-scale-yielding description to estimate an upper end value of the nucleation length (1,3) following: $L_c = \frac{\mu D_c}{\Delta\sigma_d}$. From these relationships we get a scaling between the precursory moment $M_p$, the stress drop, and the fracture energy G:

$$M_p = \frac{(2\mu G)^3}{\Delta\sigma_d^5}$$

(*Eq6*)

The seismic moment is $M_0=\mu.u_{cos}.L^2$ and the co-seismic static stress drop is $\Delta\sigma_s = C.\mu.\frac{u_{cos}}{L}$, where *C* is a geometric factor equal to 7/16 for a circular crack and $\mu$ the rock's shear modulus (*37*). From these two relationships, we get:

$$u_{cos} = \frac{\Delta\sigma_s^{\frac{2}{3}} M_0^{\frac{1}{3}}}{\mu\, C^{\frac{2}{3}}}$$

(*Eq7*)

Experimental (*2,24,25*), seismological (*26*) and theoretical studies (*27*) demonstrated that the fracture energy of earthquakes increases as a power law of their co-seismic slip, which is a proxy for the rupture length and can be written as:

$$G = a\, u_{cos}^{\alpha}$$

(*Eq8*)

Where *a* is a scaling pre-factor, and *α* a given power. In our experiments, we find a~1.22e10 and *α~1.783* (Fig.S4.).

Combining equations (*Eq6*)-(*Eq8*), we get the following scaling relation between precursory and co-seismic moments:

$$M_p = \frac{(2a.\mu^{1-\alpha})^3}{C^{2\alpha}} \cdot \frac{\Delta\sigma_s^{2\alpha}}{\Delta\sigma_d^5} \cdot M_0^{\alpha}$$

(*Eq1*)

**Acknowledgments:** M.A. and M.V. acknowledge the funding support from the Swiss National Science Foundation (SNF), project no. PYAPP2160588, and the technical staff at LEMR for help with sample preparation and Barnaby Fryer for proofreading the manuscript. M.V. acknowledges the European Research Council Starting Grant project 757290-BEFINE. A.S acknowledges the European Research Council, grant no. 681346-REALISM. F.X.P. acknowledges funding from the SNF, project no. PZENP2173613. M.A., M.V. and A.S. also acknowledge the Germaine de Staël French-Swiss exchange program.

**Author Contributions:**

**Supplementary Information:**

Supplementary Methods

Fig S1 – S5

Table S1 – S3

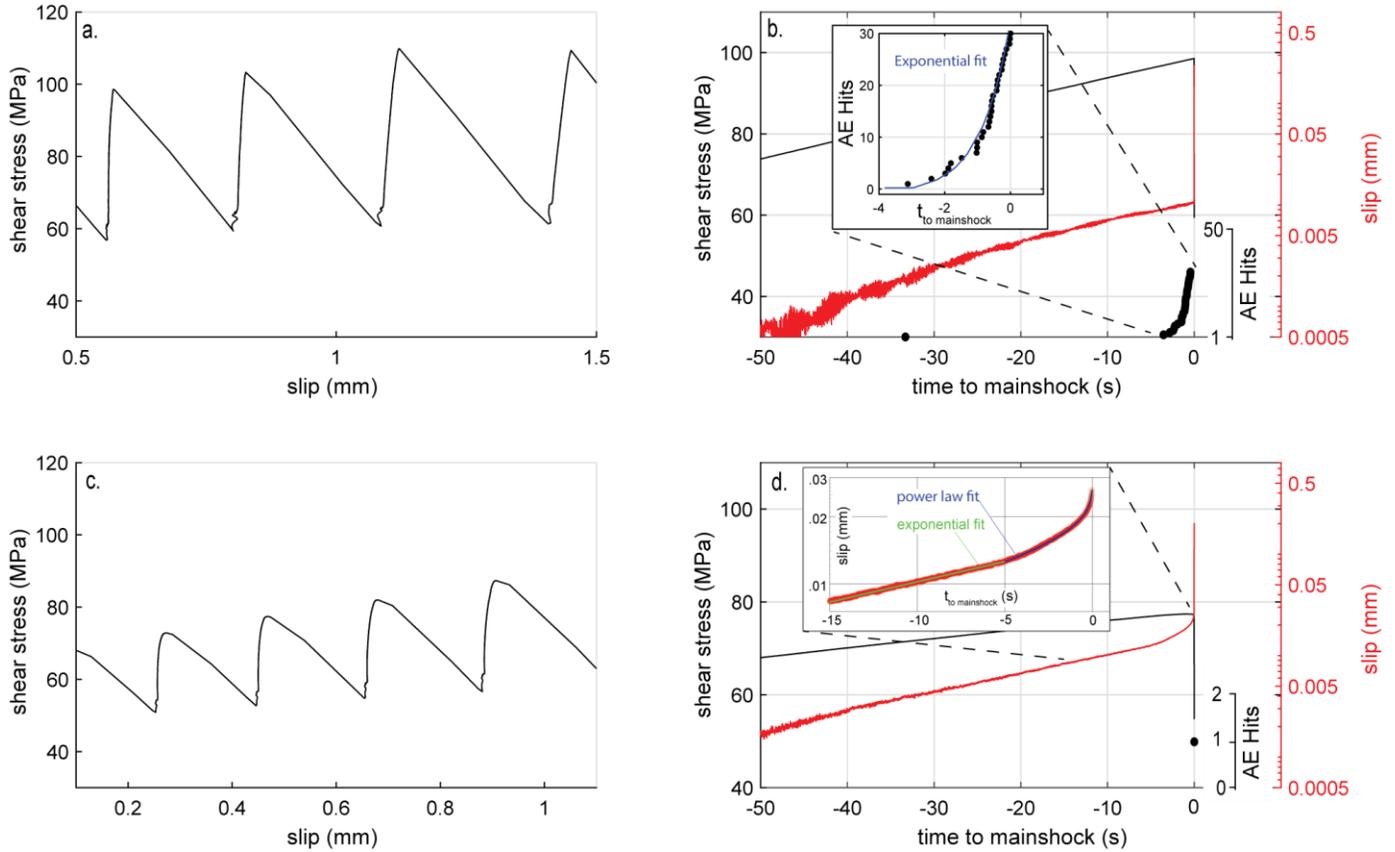

**Figure.1. Stick-slip experiments.**

**a,c.** Shear stress vs. on-fault slip for experiments run at confining pressure $\sigma_3$'=70 MPa. **a.** Dry experiment, **c.** Experiment at fluid pressure $p_f$= 1 MPa. **b,d.** Shear stress (black line), fault slip (red line in log scale) and AE's (black dots) for one stick slip cycle vs. time to mainshock ($t_0$). **b.** Dry experiment. Inset shows a close-up of AE's from $t_0$-4 s with exponential fit. **d.** Experiment at $p_f$=1 MPa. Only acoustic emission arrival was the mainshock. Inset shows exponential fit for the first nucleation stage from $t_0$-15s to $t_0$-5s as: $u(t)= 0.015 \cdot \exp(t_0-t)/23.8)$ in green. In addition, in blue, a power law fit from $t_0$-5s to $t_0$ as: $u(t)= u(t_0-5s)-0.0116*t^{0.254}$.

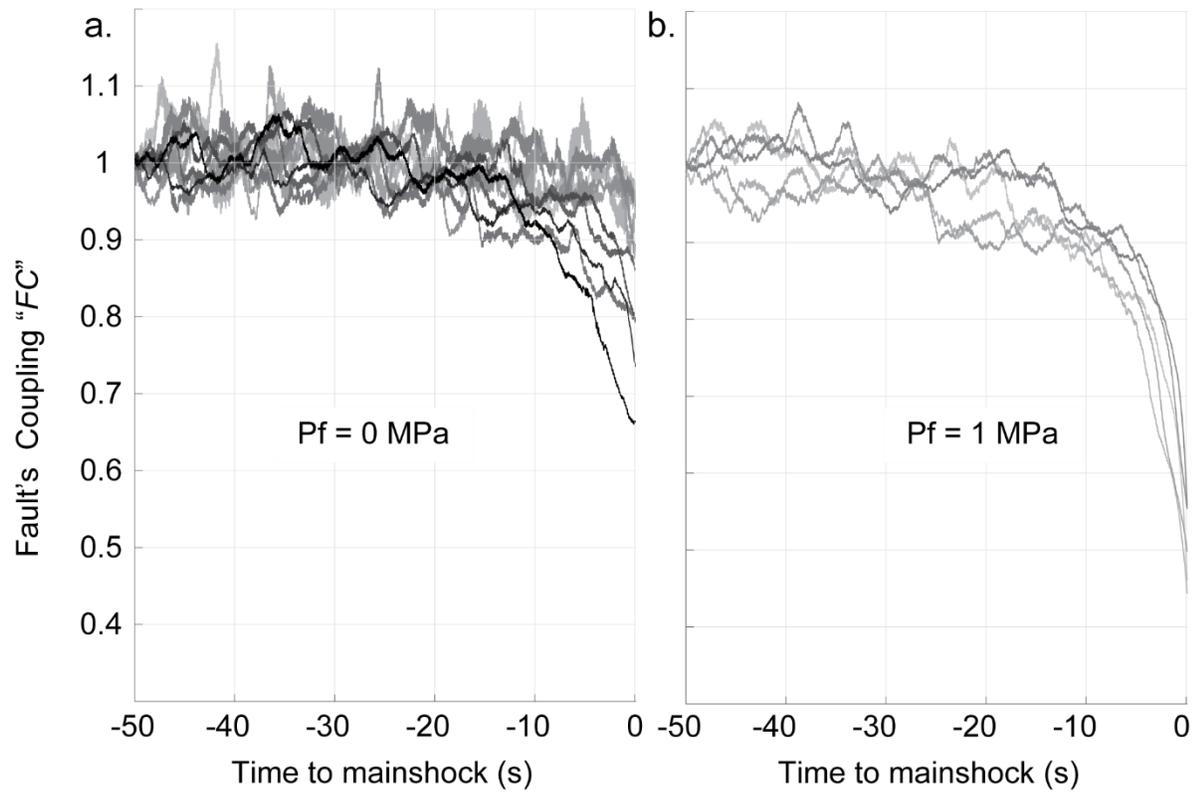

**Figure. 2. Fault coupling prior to mainshock for single stick-slip cycles.**

Fault coupling (*Methods*) vs. time to mainshock for experiments at $\sigma_3' = 70$ MPa. **a.** Dry experiments **b.** With pressurized fluid ($p_f = 1$ MPa). Darkest traces correspond to later events.

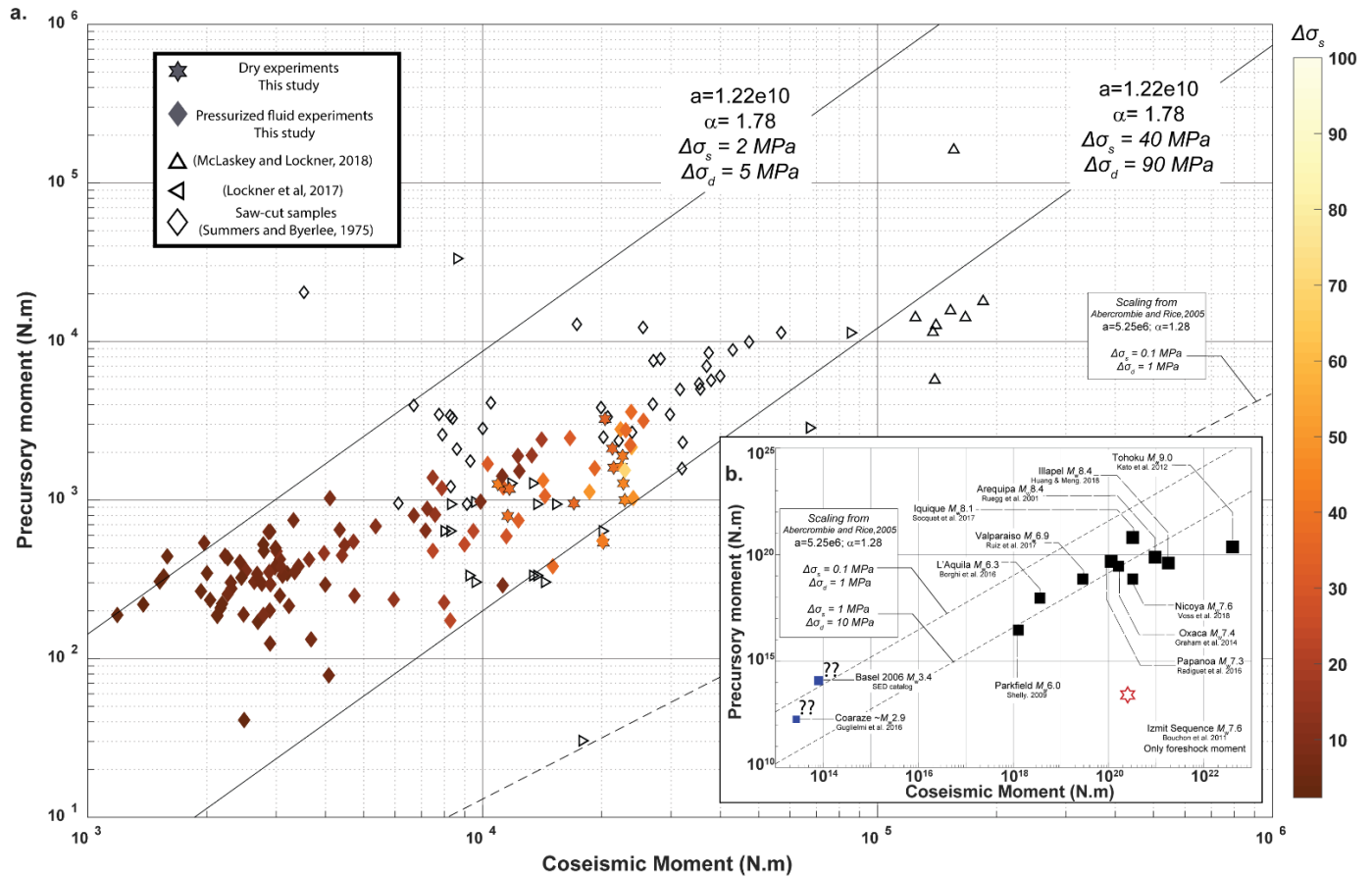

**Figure. 3. Earthquake moment release in natural and laboratory earthquakes.**
a. Precursory vs Co-seismic moment release for all stick-slip cycles analyzed from the laboratory. Color bar represents static shear stress drop. Empty symbols account for external points. Black lines account for the scaling using (Eq1) with α=1.783 and a= 1.22e10 J.m2+α. b. Data for natural earthquakes. Full squares account for Mp inferred from integrated analysis of a combination of geodetic, and/or seismic precursory moment release. Blue squares represent induced events (Methods). Red star corresponds to the Izmit sequence (18) and accounts for all 18 foreshock moment release. Black dashed lines correspond to M0 vs. Mp from Eq1 using the scaling from Ref:26 with α=1.28 and a=5.25e6.

# Supplementary Information for

## Precursory moment release scales with earthquake magnitude.

M. Acosta, F.X. Passelègue, A. Schubnel, R.Madariaga, M. Violay
Correspondence to: mateo.acosta@epfl.ch

**This PDF file includes:**

Supplementary Methods
Figs. S1 to S5
Tables S1 to S3

**Supplementary Methods**

**<u>Precursory and co-seismic moment release in natural Earthquakes.</u>**
To compute the precursory and co-seismic moment release for natural earthquakes different methods were used in the literature and are detailed below for each point of Figure.3b. The values for $M_p$ and $M_0$ are given in Table.S3. Note that in this analysis, we assume that all the recorded precursory slow slip contributes to the nucleation of the mainshock, therefore, the spatiotemporal frame taken for nucleation is the largest possible in all cases.

- 2009 $M_w$6.3 L'Aquila Earthquake from *Borghi et al. (2016)*, Ref.*38*.
Borghi et al. analyzed continuous GPS stations and detected an SSE (Slow Slip Event) that started on the 12 February and lasted for almost two weeks prior to the 2009 $M_w$6.3 L'Aquila Earthquake. From their analysis, the authors estimated a total precursory magnitude for the SSE of $M_w$5.9 which resulted in a precursory moment release, $M_p \sim 10^{\wedge}(3/2(5.9+6.07)) = 9.02e17$ N.m. For this earthquake we have $M_0 \sim 10^{\wedge}(3/2(6.3+6.07)) = 3.59e18$ N.m.

- 2011 $M_w$9.0 Tohoku-Oki Earthquake from *Kato et al. (2012)* Ref.*20*.
In their work, Kato et al. (2012) used a waveform correlation technique on the earthquake catalog preceding the 2011 $M_w$9.0 Tohoku-Oki Earthquake in order to identify migrating foreshocks towards the epicenter of the mainshock. They inferred the temporal evolution of quasi static slip of the plate interface based on small repeating earthquakes. From this analysis, the authors estimate a total aseismic moment release by slow slip transients of $M_w$ 7.1 over the area hosting foreshock distributions. This results in $M_p \sim 10^{\wedge}(3/2(7.1+6.07)) = 4.47e19$ N.m and $M_0 \sim 10^{\wedge}(3/2(9.0+6.07)) = 3.16e22$ N.m.

- 2014 $M_w$8.1 Iquique Earthquake from *Socquet et al. (2017)* Ref:*21*.
Socquet et al. (2017) analyzed the acceleration recorded at a group of GPS stations located in Coastal Northern Chile. This acceleration started 8 months prior to the 2014 $M_w$8.1 Iquique Earthquake. They showed that this acceleration corresponded to a first $M_w$6.5 slow slip event on the Chilean subduction interface, which was followed by a large $M_w$6.7 foreshock. That foreshock further generated a $M_w$7.0 afterslip event which finally ruptured the area of the Mw8.1 mainshock. From their analysis, we estimate the total precursory moment release as the sum of these events, resulting in $M_p \sim 10^{\wedge}(3/2(6.5+6.07)) + 10^{\wedge}(3/2(6.7+6.07)) + 10^{\wedge}(3/2(7.0+6.07)) = 7.16e18 + 1.43e19 + 4.03e19 = 6.17e19$ N.m. and we have $M_0 \sim 10^{\wedge}(3/2(8.1+6.07)) = 1.80e21$ N.m.

- 2004 $M_w$6.0 Parkfield Earthquake *Shelly* (*2009*), Ref:*39*
In his work, Shelly (2009) used observations of deep tremors in order to infer slow slip preceding the 2004 $M_w$6.0 Parkfield Earthquake. The author observed elevated tremor rates in the 3 months preceding the mainshock ~16km beneath the hypocenter and concluded that the deep slip interacted with the mainshock area for that earthquake. From the tremor rate analysis, the author proposes (under several assumptions) that the slow slip event prior to the mainshock had a $M_w$4.9. This results in $M_p \sim 10^{\wedge}(3/2(4.9+6.07)) = 2.85e16$ N.m and $M_0 \sim 10^{\wedge}(3/2(6.0+6.07)) = 1.27e18$ N.m.

- 2014 $M_w$7.3 Papanoa Earthquake from *Radiguet et al.* (*2016*), Ref :*40*.
In the work of Radiguet et al. (*2016*), the authors reconstructed the aseismic slip evolution on the subduction interface of the Guerrero segment through inversion of GPS position time series. The authors found a slow slip of total moment magnitude $M_w$7.6 that began two months prior to the 2014 $M_w$7.3 Papanoa earthquake. This slow slip event persisted for ~9 months after the Papanoa earthquake. The authors studied the temporal evolution of moment release during the slow slip event and found that, prior to the Papanoa earthquake, around 15% of the moment release corresponding to the $M_w$7.6 slow slip event had been released at the time of the earthquake. Using the moment magnitude scale, this results on an estimation for $M_p$~0.15*10^(3/2(7.6+6.07)) = 7.77.e19 N.m and $M_0$ ~10^(3/2(7.3+6.07)) =8.91.10e19 N.m.

- 2015 $M_w$8.4 Illapel Earthquake from *Huang and Meng.* (*2018*), Ref:*41*

Huang and Meng (*2018*), Propose that the 2014 Illapel earthquake was preceded by a progressively accelerating aseismic slip phase. Such analysis is done through a matched-filter technique which allowed for the identification of repeating earthquakes. Then, from repeating earthquake analysis, the authors propose that an area of 50.50 km$^2$ presented an average slip of ~30 cm in a time period of several months prior to the mainshock. Such an observation, assuming a rock's shear modulus of 30 GPa, results in an estimated $M_p$~30e9. (50e3)$^2$.0.3 = 2.25.e19 N.m and $M_0$ ~10^(3/2(8.4+6.07)) =5.07.10e21 N.m.

-2012 Nicoya $M_w$7.6 earthquake from *Voss et al.* (*2018*), Ref:*17*

Analyzing the slip and seismicity that preceded the 2012 Nicoya $M_w$7.6 Earthquake, Voss et al. were able to study data from 20+ GPS stations which were located in a peninsular area and therefore close to the mainshock's epicenter. From their analysis, the authors propose that a Slow Slip Event started 6 months prior to the mainshock. They conclude that the coulomb frictional stress change prior to the mainshock was low and therefore they discard a cascade-like nucleation process. The authors estimate the precursory moment release to an equivalent $M_w$6.5 event which results in an estimated $M_p$~7.16e18 N.m and $M_0$~ 3.20e19 N.m.

In addition, the authors compile a number of earthquakes where precursory aseismic slip was observed (their supplementary Table S1) from which we have plotted the additional points in figure 3 from Refs: *21,42-44*.

-2017 $M_w$6.9 Valparaiso Earthquake from *Ruiz et al.* (*2018*) Ref: *45*.

In their work, Ruiz et al. studied the nucleation phase through GPS and repeater type seismicity of the 2017 $M_w$6.9 Valparaiso earthquake. They conclude that most of the precursory deformation phase was aseismic (80% of the precursory seismic moment) and they estimate the precursory moment release to an equivalent $M_w$6.55 event which results in an estimated $M_p$~8.51e18 N.m and $M_0$~ 2.85e19 N.m.

-2006 Basel $M_w$3.2 Earthquake from SED catalog Ref:*46*

Through the precise catalog of the Swiss seismological service (SED) with a magnitude of completion $M_w$1.4, we identify all seismic events that occurred prior to the December 8$_{th}$ 2006 $M_w$3.2 induced earthquakes that occurred in Basel. Because of the low magnitude of completion and the small spatial distribution of seismic events (Haring et al., 2008), it is reasonable to assume that all the recorded events can contribute to the nucleation of the largest event. We therefore sum up the magnitudes of all recorded events prior to the mainshock and estimate $M_p$= 1.22E+14 N.m and $M_0$= 8.04E+13. (SED website reference).

-Coaraze induced earthquakes from *Guglielmi et al.* (*2015*), Ref:*28*

In the study by *Guglielmi et al., (2015)*, the authors performed semi-controlled hydraulic injections into a natural fault and recorded the fault's response in terms of slip and seismicity. From their Figure 1, a precursory aseismic slip event precedes the occurrence of micro seismicity and a slip acceleration phase. While in this study a mainshock cannot be clearly defined, we assume that the whole slip event without any seismicity contributes to precursory moment release and the second – seismic-- slip phase contributes to co-seismic moment release. From their mechanical data and modelling, the authors estimate a 15 m radius for the precursory slipping zone and a total precursory slip of 35 cm. For the 'co-seismic slip phase', the authors estimate a 35 m radius for the second slipping zone and a total slip of 95 cm. From this, considering the shear modulus of the rock to be 30 GPa, we estimate $M_p$= 1.86e12 N.m and $M_p$= 2.74e13 N.m.

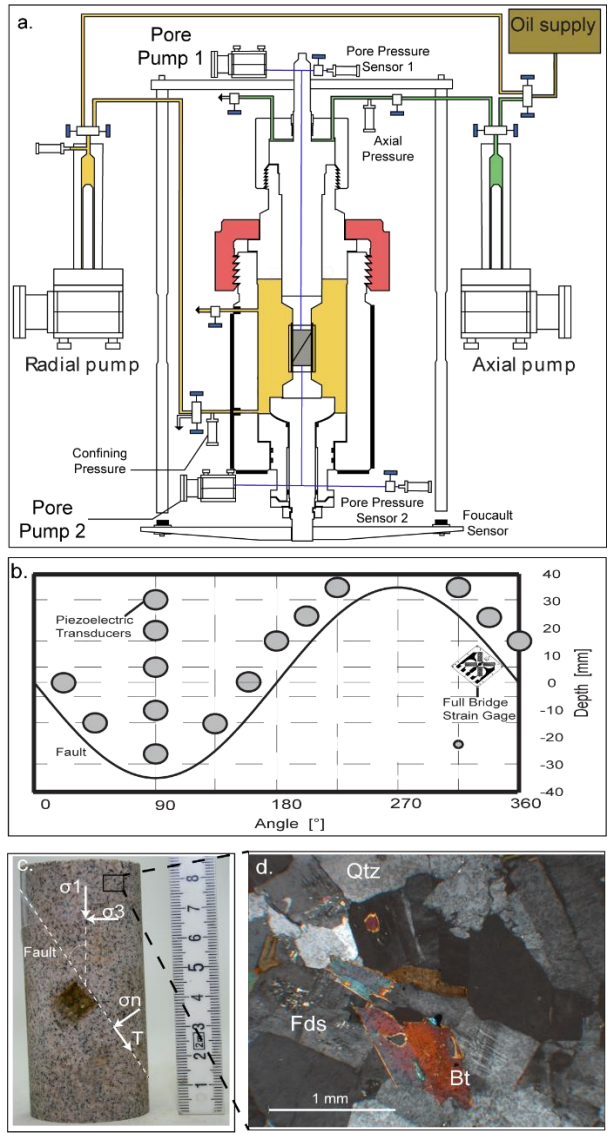

**Fig. S1. Experimental set-up.**

**a.** Triaxial apparatus. **b.** Acoustic emission sensor and strain gauge map. **c.** Westerly Granite saw-cut cylinder and stress distribution. **d.** Westerly Granite microstructure under cross-polarized optical microscopy.

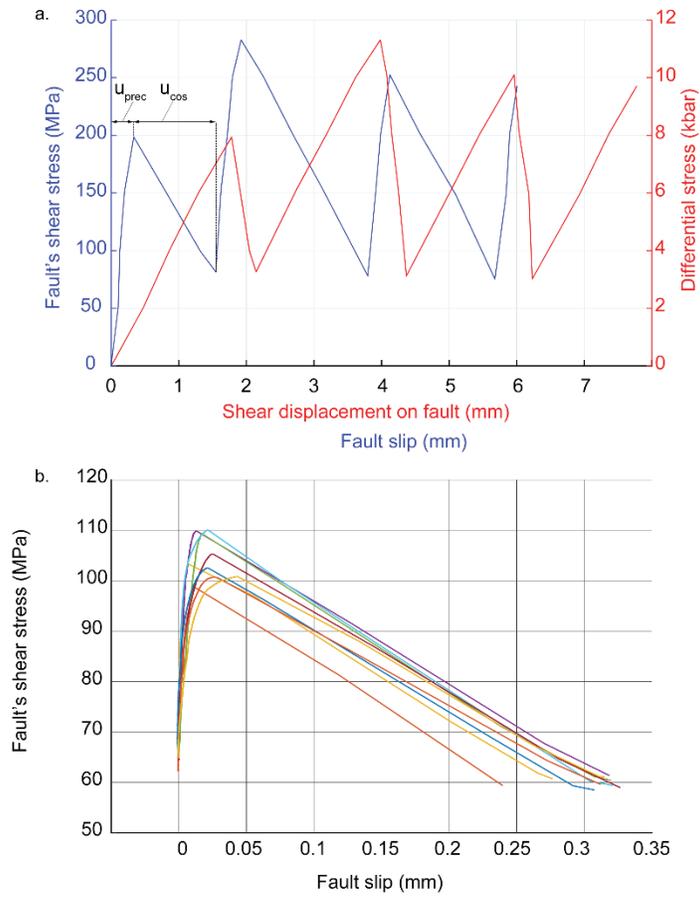

**Fig. S2. Corrections for elasticity.**

**a.** Red curve, extracted raw data for intact rock from ref (*Summers and Byerlee, 1975*). Blue curve: same dataset corrected from elastic deformation to get only fault slip and on-fault shear stress (Methods). Examples are shown of $u_{\text{prec}}$ and $u_{\text{cos}}$. values in the first event. **b.** Example of dataset from our experiments. Corrections for elasticity account for the change in shear stiffness for every event.

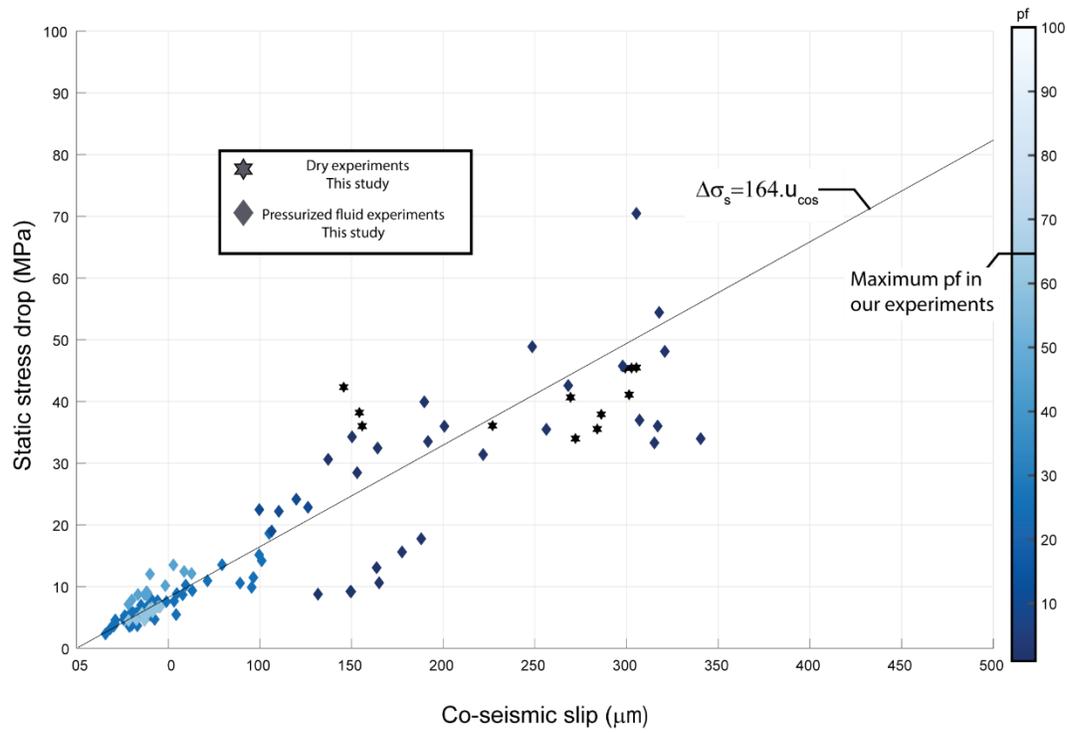

**Fig. S3. Static Stress drop versus co-seismic slip.**

Black stars represent dry experiments. Diamonds represent experiments with pressurized fluids. Pore fluid pressure in color bar. Black line represents $\Delta\sigma_s = 164 . u_{cos}$ which gives a shear modulus of 30 GPa for the fault with the circular crack approximation.

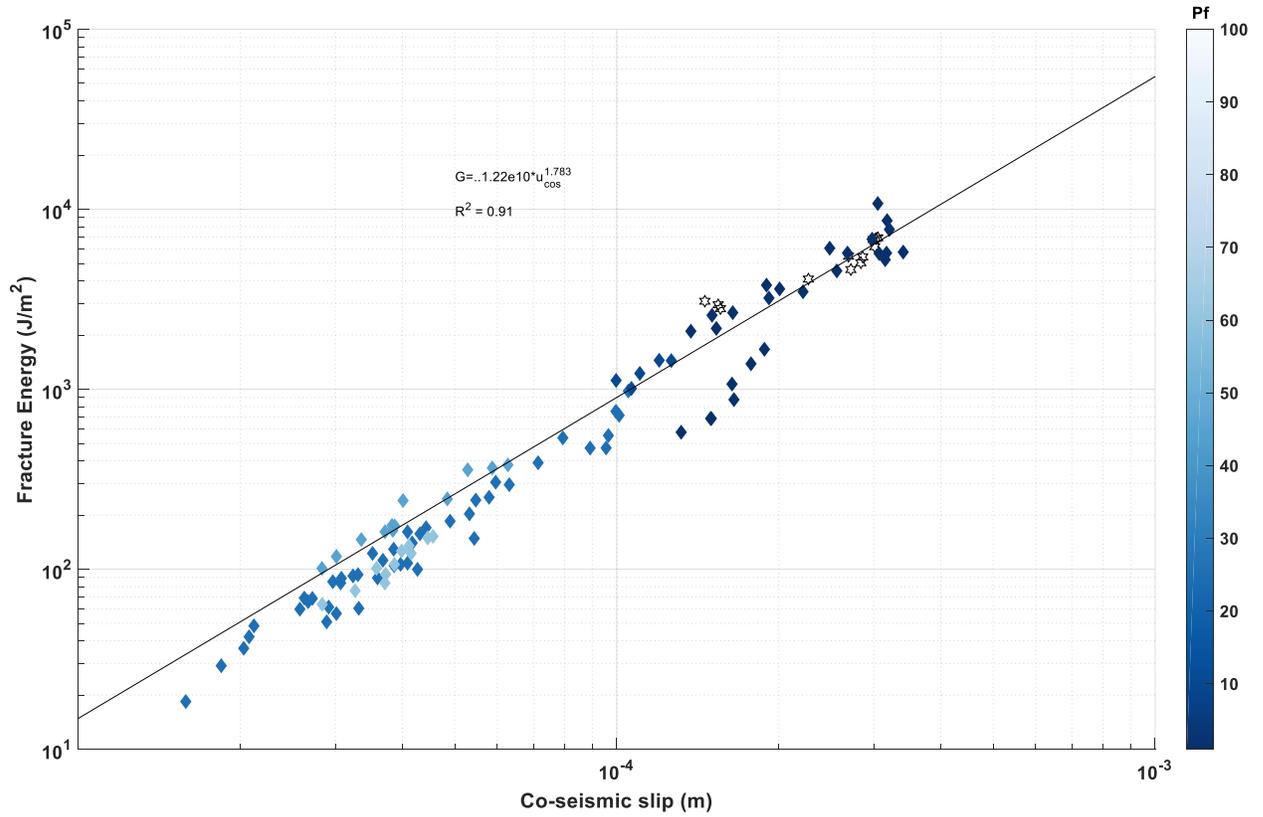

**Fig. S4. Scaling between fracture energy and co-seismic slip.**

Black stars represent dry experiments. Diamonds represent experiments with pressurized fluids. Pore fluid pressure in color bar. Black line represents $G = 1.22e10 \cdot u_{cos}^{1.783}$, which is the best fit to the data points.

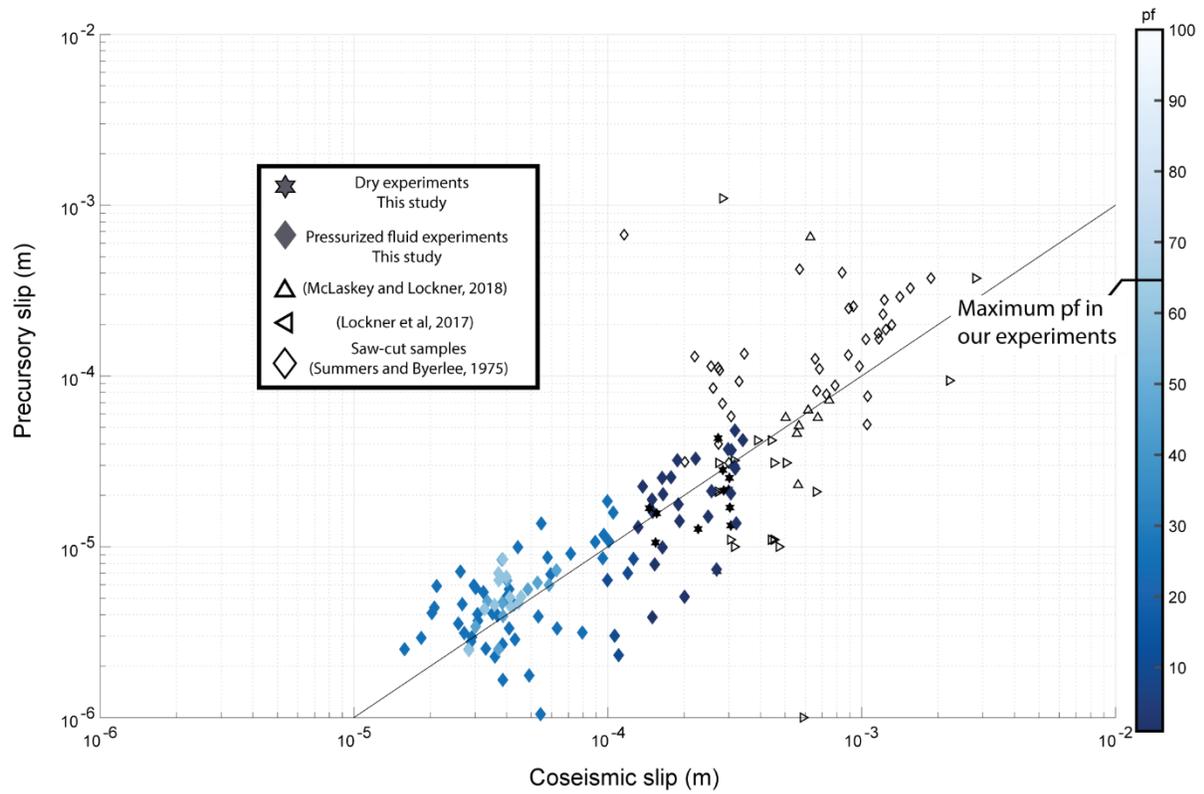

**Fig. S5. Precursory and co-seismic slip relation.**

Precursory fault slip as a function of co-seismic fault slip. Diamonds account for data from our experiments in presence of fluid pressures shown in the color bar. Black stars account for data from our dry experiments. All other symbols account for external data reported in Supplementary table 2. Black line accounts for $u_{prec}=0.1\ u_{cos}$.

**Table S1. Experimental data from this study.**

| Experiment | $\sigma_3$ | $P_f$ | $\sigma_3$' | $E_{ySystem}$ | $u_{prec}$ | $u_{cos}$ | $\Delta\sigma_s$ | $\Delta\sigma_d$ | $G'$ |
|---|---|---|---|---|---|---|---|---|---|
| Name | MPa | MPa | MPa | MPa | μm | μm | MPa | MPa | J.m$^{-2}$ |
| WG702 | 70 | 0 | 70 | 47500 | 12.7 | 226.9 | 36.07 | 76.10 | 8633.86 |
| WG703 | 70 | 0 | 70 | 45000 | 7.2 | 269.4 | 40.66 | 82.27 | 11083.38 |
| WG704 | 70 | 0 | 70 | 43500 | 13.3 | 305.3 | 45.47 | 91.16 | 13916.84 |
| WG705 | 70 | 0 | 70 | 43000 | 17.0 | 302.7 | 45.41 | 83.40 | 12623.10 |
| WG706 | 70 | 0 | 70 | 40500 | 21.6 | 299.3 | 45.38 | 91.07 | 13626.96 |
| WG707 | 70 | 0 | 70 | 40500 | 25.3 | 301.4 | 41.08 | 84.76 | 12772.22 |
| WG708 | 70 | 0 | 70 | 38500 | 21.4 | 286.2 | 37.90 | 76.28 | 10916.02 |
| WG709 | 70 | 0 | 70 | 38000 | 28.2 | 284.0 | 35.51 | 79.17 | 11242.28 |
| WG710 | 70 | 0 | 70 | 38000 | 43.3 | 272.1 | 33.98 | 73.94 | 10059.81 |
| WG5401 | 95 | 25 | 70 | 66500 | 2.8 | 28.9 | 3.53 | 14.38 | 207.99 |
| WG5402 | 95 | 25 | 70 | 72500 | 3.0 | 29.2 | 4.23 | 16.79 | 244.80 |
| WG5403 | 95 | 25 | 70 | 70500 | 5.7 | 30.1 | 3.76 | 14.71 | 221.65 |
| WG5404 | 95 | 25 | 70 | 70500 | 1.7 | 38.6 | 5.42 | 17.04 | 328.72 |
| WG5405 | 95 | 25 | 70 | 68500 | 2.7 | 38.5 | 6.71 | 18.51 | 356.43 |
| WG5406 | 95 | 25 | 70 | 68500 | 4.6 | 42.6 | 4.68 | 19.59 | 417.63 |
| WG5407 | 95 | 25 | 70 | 68500 | 4.6 | 41.7 | 6.71 | 20.52 | 427.69 |
| WG5408 | 95 | 25 | 70 | 68000 | 3.7 | 30.8 | 5.78 | 18.01 | 277.27 |
| WG5409 | 95 | 25 | 70 | 68000 | 0.5 | 33.2 | 3.66 | 20.63 | 342.12 |
| WG5410 | 95 | 25 | 70 | 68000 | 1.8 | 49.0 | 7.55 | 24.68 | 604.84 |
| WG5411 | 95 | 25 | 70 | 68000 | 8.7 | 57.9 | 8.67 | 22.92 | 663.78 |
| WG5412 | 95 | 25 | 70 | 72500 | 1.0 | 54.4 | 5.45 | 27.40 | 744.77 |
| WG5413 | 95 | 25 | 70 | 64500 | 3.1 | 79.4 | 13.53 | 34.04 | 1351.33 |
| WG5414 | 95 | 25 | 70 | 64500 | 8.6 | 95.5 | 9.87 | 38.27 | 1828.25 |
| WG5415 | 95 | 25 | 70 | 66500 | 18.5 | 99.6 | 15.13 | 41.50 | 2067.25 |
| WG5416 | 95 | 25 | 70 | 66500 | 15.8 | 105.1 | 18.60 | 38.81 | 2038.57 |
| WG5501 | 95 | 1 | 94 | 80000 | 22.5 | 137.3 | 30.60 | 73.52 | 5046.04 |
| WG5502 | 95 | 1 | 94 | 77000 | 16.0 | 150.3 | 34.25 | 85.67 | 6437.83 |
| WG5503 | 95 | 1 | 94 | 74000 | 17.7 | 189.7 | 39.92 | 87.68 | 8316.70 |
| WG5504 | 95 | 1 | 94 | 66000 | 15.0 | 248.6 | 48.86 | 95.81 | 11907.25 |
| WG5505 | 95 | 1 | 94 | 64000 | 20.5 | 305.4 | 70.45 | 103.48 | 15798.84 |
| WG5506 | 95 | 1 | 94 | 60000 | 13.7 | 320.9 | 48.10 | 98.09 | 15736.73 |
| WG5507 | 95 | 1 | 94 | 58000 | 7.4 | 268.2 | 42.56 | 94.62 | 12688.52 |
| WG5508 | 95 | 1 | 94 | 56000 | 28.7 | 317.8 | 54.41 | 92.98 | 14772.78 |
| WG5509 | 95 | 1 | 94 | 56000 | 37.3 | 297.9 | 45.73 | 83.84 | 12488.74 |
| WG9401 | 70 | 0 | 70 | 57000 | 16.8 | 145.7 | 42.30 | 84.62 | 6165.93 |
| WG9402 | 70 | 0 | 70 | 54500 | 10.6 | 154.3 | 38.19 | 84.91 | 6549.57 |

| Experiment | $\sigma_3$ | $Pf$ | $\sigma_3'$ | $E_{ySystem}$ | $u_{prec}$ | $u_{cos}$ | $\Delta\sigma_s$ | $\Delta\sigma_d$ | $G'$ |
|---|---|---|---|---|---|---|---|---|---|
| Name | MPa | MPa | MPa | MPa | μm | μm | MPa | MPa | J.m$^{-2}$ |
| WG9403 | 70 | 0 | 70 | 57000 | 15.7 | 155.8 | 36.01 | 74.49 | 5802.90 |
| WG10101 | 95 | 25 | 70 | 40000 | 2.5 | 15.8 | 2.32 | 14.34 | 113.55 |
| WG10102 | 95 | 25 | 70 | 37000 | 2.9 | 18.4 | 3.16 | 16.51 | 152.04 |
| WG10103 | 95 | 25 | 70 | 40000 | 4.1 | 20.3 | 3.58 | 17.95 | 182.00 |
| WG10104 | 95 | 25 | 70 | 41000 | 4.4 | 20.8 | 4.06 | 18.73 | 194.40 |
| WG10105 | 95 | 25 | 70 | 41000 | 5.9 | 21.2 | 4.58 | 19.09 | 202.27 |
| WG10106 | 95 | 25 | 70 | 37000 | 3.6 | 25.8 | 4.65 | 17.06 | 220.01 |
| WG10107 | 95 | 25 | 70 | 40000 | 4.6 | 26.7 | 4.95 | 16.90 | 225.73 |
| WG10108 | 95 | 25 | 70 | 40000 | 3.1 | 27.2 | 5.06 | 19.36 | 263.28 |
| WG10109 | 95 | 25 | 70 | 40000 | 7.2 | 26.3 | 5.26 | 19.19 | 252.19 |
| WG10110 | 95 | 25 | 70 | 40000 | 4.0 | 30.7 | 5.46 | 23.22 | 356.41 |
| WG10111 | 95 | 25 | 70 | 38000 | 2.5 | 33.1 | 5.64 | 19.13 | 316.20 |
| WG10112 | 95 | 25 | 70 | 40000 | 5.9 | 29.7 | 5.74 | 22.96 | 340.80 |
| WG10113 | 95 | 25 | 70 | 40000 | 5.4 | 32.4 | 5.67 | 24.24 | 392.22 |
| WG10114 | 95 | 25 | 70 | 40000 | 2.3 | 35.9 | 4.98 | 20.21 | 363.16 |
| WG10115 | 95 | 25 | 70 | 40000 | 4.1 | 35.2 | 6.96 | 20.37 | 358.39 |
| WG10116 | 95 | 25 | 70 | 40000 | 4.0 | 36.8 | 6.09 | 21.68 | 398.71 |
| WG10117 | 95 | 25 | 70 | 40000 | 3.3 | 40.9 | 5.29 | 27.22 | 556.10 |
| WG10118 | 95 | 25 | 70 | 41000 | 2.9 | 43.1 | 7.30 | 22.43 | 483.34 |
| WG10119 | 95 | 25 | 70 | 43000 | 5.6 | 40.9 | 7.89 | 22.53 | 460.42 |
| WG10120 | 95 | 25 | 70 | 43000 | 5.0 | 39.7 | 5.37 | 22.21 | 440.43 |
| WG10121 | 95 | 25 | 70 | 43000 | 3.3 | 63.1 | 9.34 | 29.18 | 921.37 |
| WG10122 | 95 | 25 | 70 | 43000 | 3.9 | 53.3 | 7.62 | 27.41 | 729.87 |
| WG10123 | 95 | 25 | 70 | 45000 | 10.0 | 44.2 | 7.69 | 27.41 | 606.35 |
| WG10124 | 95 | 25 | 70 | 45000 | 6.9 | 59.6 | 10.22 | 23.22 | 691.69 |
| WG10125 | 95 | 25 | 70 | 52000 | 13.7 | 54.7 | 8.85 | 28.59 | 782.21 |
| WG10126 | 95 | 25 | 70 | 47000 | 9.1 | 71.4 | 10.93 | 35.47 | 1266.82 |
| WG10127 | 95 | 25 | 70 | 47000 | 10.7 | 89.2 | 10.56 | 33.18 | 1480.05 |
| WG10128 | 95 | 25 | 70 | 47000 | 11.8 | 96.5 | 11.46 | 31.51 | 1519.70 |
| WG10129 | 95 | 25 | 70 | 47000 | 10.8 | 101.0 | 14.18 | 34.19 | 1726.81 |
| WG10201 | 95 | 45 | 50 | 50000 | 3.4 | 30.1 | 7.81 | 23.56 | 355.18 |
| WG10202 | 95 | 45 | 50 | 58000 | 2.5 | 28.3 | 7.13 | 21.44 | 303.93 |
| WG10203 | 95 | 45 | 50 | 53000 | 4.8 | 33.5 | 8.71 | 22.97 | 385.21 |
| WG10204 | 95 | 45 | 50 | 54000 | 2.5 | 37.1 | 8.70 | 27.07 | 502.30 |
| WG10205 | 95 | 45 | 50 | 54000 | 3.9 | 38.7 | 9.01 | 25.92 | 501.22 |
| WG10206 | 95 | 45 | 50 | 54000 | 6.3 | 40.1 | 12.01 | 26.75 | 536.22 |
| WG10207 | 95 | 45 | 50 | 59000 | 4.7 | 38.4 | 8.57 | 33.21 | 637.13 |
| WG10208 | 95 | 45 | 50 | 61000 | 8.4 | 38.2 | 9.15 | 26.37 | 504.27 |

| Experiment | $\sigma_3$ | $P_f$ | $\sigma_3'$ | $E_{ySystem}$ | $u_{prec}$ | $u_{cos}$ | $\Delta\sigma_s$ | $\Delta\sigma_d$ | $G'$ |
|---|---|---|---|---|---|---|---|---|---|
| Name | MPa | MPa | MPa | MPa | μm | μm | MPa | MPa | J.m$^{-2}$ |
| WG10209 | 95 | 45 | 50 | 53000 | 5.6 | 48.4 | 10.14 | 31.61 | 765.73 |
| WG10210 | 95 | 45 | 50 | 53000 | 6.2 | 52.9 | 13.49 | 28.52 | 753.84 |
| WG10211 | 95 | 45 | 50 | 51000 | 6.0 | 58.7 | 12.44 | 29.88 | 877.07 |
| WG10212 | 95 | 45 | 50 | 50000 | 7.3 | 62.8 | 12.10 | 36.16 | 1134.83 |
| WG10501 | 95 | 60 | 35 | 37000 | 2.5 | 28.4 | 4.50 | 18.04 | 255.90 |
| WG10502 | 95 | 60 | 35 | 42000 | 4.3 | 32.7 | 4.65 | 15.65 | 255.54 |
| WG10503 | 95 | 60 | 35 | 39000 | 4.6 | 35.8 | 5.65 | 21.81 | 390.32 |
| WG10504 | 95 | 60 | 35 | 41000 | 6.4 | 37.2 | 5.04 | 18.47 | 343.55 |
| WG10505 | 95 | 60 | 35 | 40000 | 7.0 | 37.1 | 4.52 | 27.22 | 504.94 |
| WG10506 | 95 | 60 | 35 | 41000 | 8.4 | 38.6 | 5.50 | 16.01 | 309.18 |
| WG10507 | 95 | 60 | 35 | 39000 | 4.5 | 41.5 | 5.91 | 23.35 | 484.62 |
| WG10508 | 95 | 60 | 35 | 43000 | 6.7 | 39.8 | 6.34 | 23.25 | 463.26 |
| WG10509 | 95 | 60 | 35 | 40000 | 5.0 | 41.1 | 6.56 | 19.45 | 399.94 |
| WG10510 | 95 | 60 | 35 | 37000 | 5.1 | 45.6 | 6.69 | 21.02 | 478.97 |
| WG10511 | 95 | 60 | 35 | 40000 | 4.7 | 44.5 | 6.70 | 21.75 | 484.31 |
| WG10601 | 95 | 10 | 85 | 56500 | 6.4 | 99.7 | 22.46 | 52.96 | 2640.43 |
| WG10602 | 95 | 10 | 85 | 56500 | 3.0 | 106.5 | 18.98 | 60.75 | 3233.94 |
| WG10603 | 95 | 10 | 85 | 55500 | 2.3 | 110.3 | 22.19 | 59.21 | 3267.07 |
| WG10604 | 95 | 10 | 85 | 55500 | 7.0 | 119.9 | 24.14 | 56.96 | 3414.20 |
| WG10605 | 95 | 10 | 85 | 55500 | 8.5 | 126.3 | 22.83 | 51.02 | 3220.82 |
| WG10701 | 95 | 1 | 94 | 51500 | 7.9 | 153.1 | 28.45 | 113.58 | 8693.15 |
| WG10702 | 95 | 1 | 94 | 50500 | 9.9 | 164.2 | 32.44 | 77.04 | 6325.86 |
| WG10703 | 95 | 1 | 94 | 48500 | 14.1 | 191.7 | 33.50 | 114.68 | 10992.15 |
| WG10704 | 95 | 1 | 94 | 46000 | 5.1 | 200.6 | 35.97 | 93.77 | 9406.92 |
| WG17101 | 51 | 1 | 50 | 34500 | 3.9 | 149.8 | 9.17 | 96.79 | 7250.74 |
| WG17102 | 51 | 1 | 50 | 36000 | 13.0 | 131.7 | 8.76 | 78.64 | 5178.75 |
| WG17103 | 51 | 1 | 50 | 36000 | 18.9 | 149.4 | 9.20 | 89.73 | 6702.57 |
| WG17201 | 71 | 1 | 70 | 34000 | 20.3 | 165.1 | 10.60 | 59.61 | 4920.05 |
| WG17202 | 71 | 1 | 70 | 38500 | 25.3 | 163.7 | 13.05 | 104.30 | 8536.63 |
| WG17203 | 71 | 1 | 70 | 38500 | 25.5 | 177.6 | 15.60 | 78.68 | 6985.21 |
| WG17204 | 71 | 1 | 70 | 40000 | 32.1 | 188.0 | 17.74 | 93.73 | 8811.08 |
| WG17205 | 71 | 1 | 70 | 39500 | 32.8 | 221.8 | 31.38 | 116.85 | 12956.90 |
| WG17301 | 95 | 1 | 94 | 38500 | 21.2 | 256.3 | 35.47 | 90.86 | 11641.93 |
| WG17302 | 95 | 1 | 94 | 41000 | 36.8 | 307.1 | 36.96 | 115.99 | 17809.10 |
| WG17303 | 95 | 1 | 94 | 41000 | 29.7 | 315.2 | 33.28 | 164.32 | 25895.63 |
| WG17304 | 95 | 1 | 94 | 42000 | 42.1 | 340.4 | 33.96 | 119.86 | 20401.38 |
| WG17305 | 95 | 1 | 94 | 42500 | 48.0 | 317.0 | 36.01 | 119.86 | 18995.91 |

**Table S2. External Data from Laboratory Earthquakes.**

| Study | Event | $\sigma_3$ | Diameter | $u_{prec}$ | $u_{cos}$ | $\Delta\sigma_s$ |
|---|---|---|---|---|---|---|
| Reference | Name | MPa | m | μm | μm | MPa |
| Lockner et al 2017 | Event1 | 100 | 0.0254 | 1097 | 283 | 44.37 |
| Lockner et al 2017 | Event2 | 100 | 0.0254 | 42 | 388 | 53.62 |
| Lockner et al 2017 | Event3 | 100 | 0.0254 | 21 | 661 | 93.47 |
| Lockner et al 2017 | Event4 | 100 | 0.0254 | 1 | 587 | 80.79 |
| Lockner et al 2017 | Event5 | 100 | 0.0254 | 11 | 304 | 45.56 |
| Lockner et al 2017 | Event6 | 100 | 0.0254 | 10 | 472 | 73.79 |
| Lockner et al 2017 | Event7 | 100 | 0.0254 | 42 | 440 | 68.64 |
| Lockner et al 2017 | Event8 | 100 | 0.0254 | 11 | 451 | 71.54 |
| Lockner et al 2017 | Event9 | 100 | 0.0254 | 31 | 273 | 43.82 |
| Lockner et al 2017 | Event10 | 100 | 0.0254 | 11 | 440 | 71.56 |
| Lockner et al 2017 | Event11 | 100 | 0.0254 | 31 | 504 | 81.95 |
| Lockner et al 2017 | Event12 | 100 | 0.0254 | 10 | 315 | 53.61 |
| Lockner et al 2017 | Event13 | 100 | 0.0254 | 31 | 451 | 74.94 |
| Lockner et al 2017 | Event14 | 100 | 0.0254 | 32 | 314 | 50.8 |
| Lockner et al 2017 | Event15 | 100 | 0.0254 | 21 | 263 | 48.98 |
| Lockner et al 2017 | Event16 | 100 | 0.0254 | 21 | 272 | 47.85 |
| Lockner et al 2017 | Event1 | 100 | 0.0254 | 373.4 | 2810.6 | 343.248 |
| Lockner et al 2017 | Event2 | 100 | 0.0254 | 94 | 2213 | 286.648 |
| McLaskey & Lockner 2018 | DSC1 | 80 | 0.0726 | 650 | 628 | 30.94 |
| McLaskey & Lockner 2018 | DSC2 | 80 | 0.0726 | 46 | 556 | 31.91 |
| McLaskey & Lockner 2018 | DSC3 | 80 | 0.0726 | 72 | 745 | 43.74 |
| McLaskey & Lockner 2018 | DSC4 | 80 | 0.0726 | 57 | 502 | 28.23 |
| McLaskey & Lockner 2018 | DSC5 | 80 | 0.0726 | 63 | 616 | 36.93 |
| McLaskey & Lockner 2018 | DSC6 | 80 | 0.0726 | 51 | 566 | 33.98 |
| McLaskey & Lockner 2018 | DSC7 | 80 | 0.0726 | 57 | 672 | 40.34 |
| McLaskey & Lockner 2018 | DSC8 | 80 | 0.0726 | 23 | 562 | 33.24 |
| Summers Byerlee 1977 | Fig85-Evt1 | 398 | 0.0254 | 671.4 | 116 | 9.29 |
| Summers Byerlee 1977 | Fig85-Evt2 | 398 | 0.0254 | 31.4 | 201.2 | 14.28 |
| Summers Byerlee 1977 | Fig85-Evt3 | 398 | 0.0254 | 69 | 283 | 22.14 |
| Summers Byerlee 1977 | Fig85-Evt4 | 398 | 0.0254 | 31 | 300 | 27.85 |
| Summers Byerlee 1977 | Fig85-Evt5 | 398 | 0.0254 | 40 | 273 | 24.28 |
| Summers Byerlee 1977 | Fig85-Evt6 | 398 | 0.0254 | 108 | 276 | 27.84 |
| Summers Byerlee 1977 | Fig85-Evt7 | 398 | 0.0254 | 93 | 329 | 28.56 |
| Summers Byerlee 1977 | Fig85-Evt8 | 398 | 0.0254 | 112 | 272 | 30.71 |
| Summers Byerlee 1977 | Fig85-Evt9 | 398 | 0.0254 | 135 | 345 | 30.7 |
| Summers Byerlee 1977 | Fig85-Evt10 | 398 | 0.0254 | 85 | 260 | 25.71 |

| Study | Event | $\sigma_3$ | Diameter | $u_{prec}$ | $u_{cos}$ | $\Delta\sigma_s$ |
|---|---|---|---|---|---|---|
| Reference | Name | MPa | m | μm | μm | MPa |
| Summers Byerlee 1977 | Fig85-Evt11 | 398 | 0.0254 | 114 | 255 | 24.99 |
| Summers Byerlee 1977 | Fig85-Evt12 | 398 | 0.0254 | 130 | 220 | 23.56 |
| Summers Byerlee 1977 | Fig85-Evt13 | 398 | 0.0254 | 58 | 306 | 22.14 |
| Summers Byerlee 1977 | Fig85-Evt1 | 320 | 0.0254 | 421.9 | 569.6 | 64.93 |
| Summers Byerlee 1977 | Fig85-Evt2 | 320 | 0.0254 | 132.5 | 887 | 80.53 |
| Summers Byerlee 1977 | Fig85-Evt3 | 320 | 0.0254 | 110 | 682 | 77.07 |
| Summers Byerlee 1977 | Fig85-Evt4 | 320 | 0.0254 | 88 | 786 | 91.37 |
| Summers Byerlee 1977 | Fig85-Evt5 | 320 | 0.0254 | 78 | 727 | 79.96 |
| Summers Byerlee 1977 | Fig85-Evt6 | 320 | 0.0254 | 126 | 656 | 73.57 |
| Summers Byerlee 1977 | Fig85-Evt7 | 320 | 0.0254 | 82 | 665 | 71.36 |
| Summers Byerlee 1977 | Fig85-Evt1 | 240 | 0.0254 | 248.9 | 890.1 | 95.7 |
| Summers Byerlee 1977 | Fig85-Evt2 | 240 | 0.0254 | 76 | 1056 | 118.55 |
| Summers Byerlee 1977 | Fig85-Evt3 | 240 | 0.0254 | 199 | 1314 | 150.7 |
| Summers Byerlee 1977 | Fig85-Evt4 | 240 | 0.0254 | 165 | 1170 | 134.2 |
| Summers Byerlee 1977 | Fig85-Evt1 | 78 | 0.0254 | 403.2 | 837.8 | 98.52 |
| Summers Byerlee 1977 | Fig85-Evt2 | 78 | 0.0254 | 164 | 1039 | 122.79 |
| Summers Byerlee 1977 | Fig85-Evt3 | 78 | 0.0254 | 178 | 1162 | 136.4 |
| Summers Byerlee 1977 | Fig85-Evt4 | 78 | 0.0254 | 52 | 1051 | 117.84 |
| Summers Byerlee 1977 | Fig85-Evt5 | 78 | 0.0254 | 114 | 980 | 111.41 |
| Summers Byerlee 1977 | Fig86-Evt1 | 547 | 0.0254 | 255.7 | 928.3 | 104.44 |
| Summers Byerlee 1977 | Fig86-Evt2 | 547 | 0.0254 | 279 | 1229 | 145.73 |
| Summers Byerlee 1977 | Fig86-Evt3 | 547 | 0.0254 | 291 | 1414 | 161.23 |
| Summers Byerlee 1977 | Fig86-Evt4 | 547 | 0.0254 | 187 | 1246 | 138.03 |
| Summers Byerlee 1977 | Fig86-Evt1 | 630 | 0.0254 | 230 | 1214 | 117.02 |
| Summers Byerlee 1977 | Fig86-Evt2 | 630 | 0.0254 | 374 | 1875 | 204.62 |
| Summers Byerlee 1977 | Fig86-Evt3 | 630 | 0.0254 | 327 | 1555 | 177.32 |
| Summers Byerlee 1977 | Fig86-Evt1 | 547 | 0.0254 | 291 | 1414 | 161.23 |
| Summers Byerlee 1977 | Fig86-Evt2 | 547 | 0.0254 | 187 | 1246 | 138.03 |
| Summers Byerlee 1977 | Fig86-Evt1 | 630 | 0.0254 | 230 | 1214 | 117.02 |
| Summers Byerlee 1977 | Fig86-Evt2 | 630 | 0.0254 | 374 | 1875 | 204.62 |
| Summers Byerlee 1977 | Fig86-Evt3 | 630 | 0.0254 | 327 | 1555 | 177.32 |

**Table S3. Data from Natural Earthquakes.**

| Earthquake | Detection Method | Reference | Coseismic Magnitude | Precursory Magnitude | Coseismic Moment | Precursory Moment |
|---|---|---|---|---|---|---|
| | | | Mw | Mw | N.m | N.m |
| 2009 L'Aquila, IT | GPS | Borghi et al., 2016 | 6.30 | 5.90 | 3.59E+18 | 9.02E+17 |
| 2011 Tohoku-Oki, JP | Repeaters +GPS | Kato et al., 2012 | 9.00 | 7.03 | 3.16E+22 | 4.47E+19 |
| 2014 Iquique, CL | GPS | Socquet et al., 2017 | 8.10 | 7.12 | 1.80E+21 | 6.17E+19 |
| 2014 Papanoa, MX | GPS | Radiguet et al., 2016 | 7.30 | 6.98 | 1.14E+20 | 3.78E+19 |
| 2015 Illapel, CL | Repeaters | Huang & Meng., 2018 | 8.40 | 6.83 | 5.07E+21 | 2.25E+19 |
| 2017 Valparaiso, CL | Repeaters +GPS | Ruiz et al., 2017 | 6.90 | 6.50 | 2.85E+19 | 7.16E+18 |
| 2012 Oxaca, MX | GPS | Graham et al., 2014 | 7.40 | 6.90 | 1.60E+20 | 2.85E+19 |
| 2001 Arequipa, PE | GPS | Ruegg et al., 2011 | 7.60 | 7.80 | 3.20E+20 | 6.38E+20 |
| 2012 Nicoya, CR | GPS | Voss et al., 2018 | 7.60 | 6.50 | 3.20E+20 | 2.81E+21 |
| 2004 Parkfield, USA | Deep Tremor | Shelly, 2009 | 6.00 | 4.90 | 1.27E+18 | 2.85E+16 |